\documentclass[twocolumn,10pt]{article}

\usepackage[margin=0.85in,columnsep=0.28in]{geometry}
\usepackage{mathptmx}
\usepackage[T1]{fontenc}
\usepackage{amsmath, amssymb}
\usepackage{booktabs}
\usepackage{graphicx}
\usepackage{xcolor}
\usepackage{enumitem}
\usepackage[colorlinks=true,linkcolor=blue!60!black,citecolor=blue!60!black,urlcolor=blue!60!black]{hyperref}
\setlist{nosep}

\begin{document}

\twocolumn[{%
  \centering
  {\LARGE\bfseries TraceCompiler: Skill-Guided Mining and Compilation of
   LLM Agent Traces into Mostly Deterministic Workflows\par}
  \vspace{1.2em}
  {\large Salma El Yadouni\textsuperscript{1} \quad
          Guanyi Li\textsuperscript{2}\par}
  \vspace{0.45em}
  {\normalsize \textsuperscript{1}EPFL, Lausanne, Switzerland \quad
               \textsuperscript{2}Binome Technologies, London, UK\par}
  \vspace{0.3em}
  {\small \texttt{salma.elyadouni@epfl.ch} \quad \texttt{craig@binome.dev}\par}
  \vspace{1.6em}
  \begin{minipage}{0.88\textwidth}
  \small
  \textbf{Abstract.}
  Tool-using language-model agents repeatedly rediscover procedures they have
  already executed, producing traces that mix reusable structure with retries,
  exploration, accidental ordering, and repeated lookups. We present
  \textbf{TraceCompiler}, a skill-guided system that mines clusters of noisy
  agent traces and compiles them into executable, mostly deterministic
  workflows. It admits an inter-tool dependency only when a consumer argument
  contains a value attributable uniquely to an earlier producer; every hard edge
  carries an auditable evidence tuple, and ambiguous relations are marked
  \emph{suspected} and impose no ordering constraint. Bindings are classified as
  constants, user inputs, copied outputs, transforms, or residual LLM decisions.
  On T1, a mechanized form of the rule recovers producer--consumer dependencies
  at 0.928 precision and 0.943 recall over 15{,}775 def--use edges of its
  training split, against 0.711 $F_1$ for adjacency and 0.712 for a
  frequency-thresholded directly-follows measure on identical data; the compiler
  skill run blind reaches 0.992 on 250 of those edges. On AppWorld we replay
  released trajectories in the deterministic simulator to recover masked return
  values and measure the rule against 563 token edges at 0.993 precision --- a
  self-consistency check, since replay injects tokens by a related heuristic. We
  compile two recurring intents: a Venmo money-request intent reduces 34
  observed API calls to 11 runtime calls and, under leave-one-out execution
  against the benchmark's own state tests, passes 15 of 21, the failing fold
  escalating rather than acting because its required branch was never observed;
  and a Spotify/Todoist intent the compiler correctly refuses to compile,
  because an irreversible side effect is under-determined. We measure call
  reduction but not offline compilation cost, so we claim no net efficiency
  result.
  \end{minipage}
  \vspace{2em}
}]

\section{Introduction}
\label{sec:intro}

Tool-using language-model agents repeatedly pay to rediscover procedures they
have already executed. A recurring intent --- listing open tickets, requesting a
payment --- is issued again and again, and on each occurrence the agent
re-derives the same tool sequence: it re-resolves stable identity and
configuration, retries equivalent calls with modified parameters, reads tool
documentation it has read before, and carries an expanding reasoning history.
Most of this is not intrinsic to the task; it is the cost of treating every
request as new. Recent serving-time work reinforces the premise, profitably
predicting agents' upcoming tool calls within a
session~\cite{paste,speculativeactions}: the regularity such systems exploit
transiently is what compilation can capture persistently.

\begin{figure*}[t]
\centering
\includegraphics[width=0.86\textwidth]{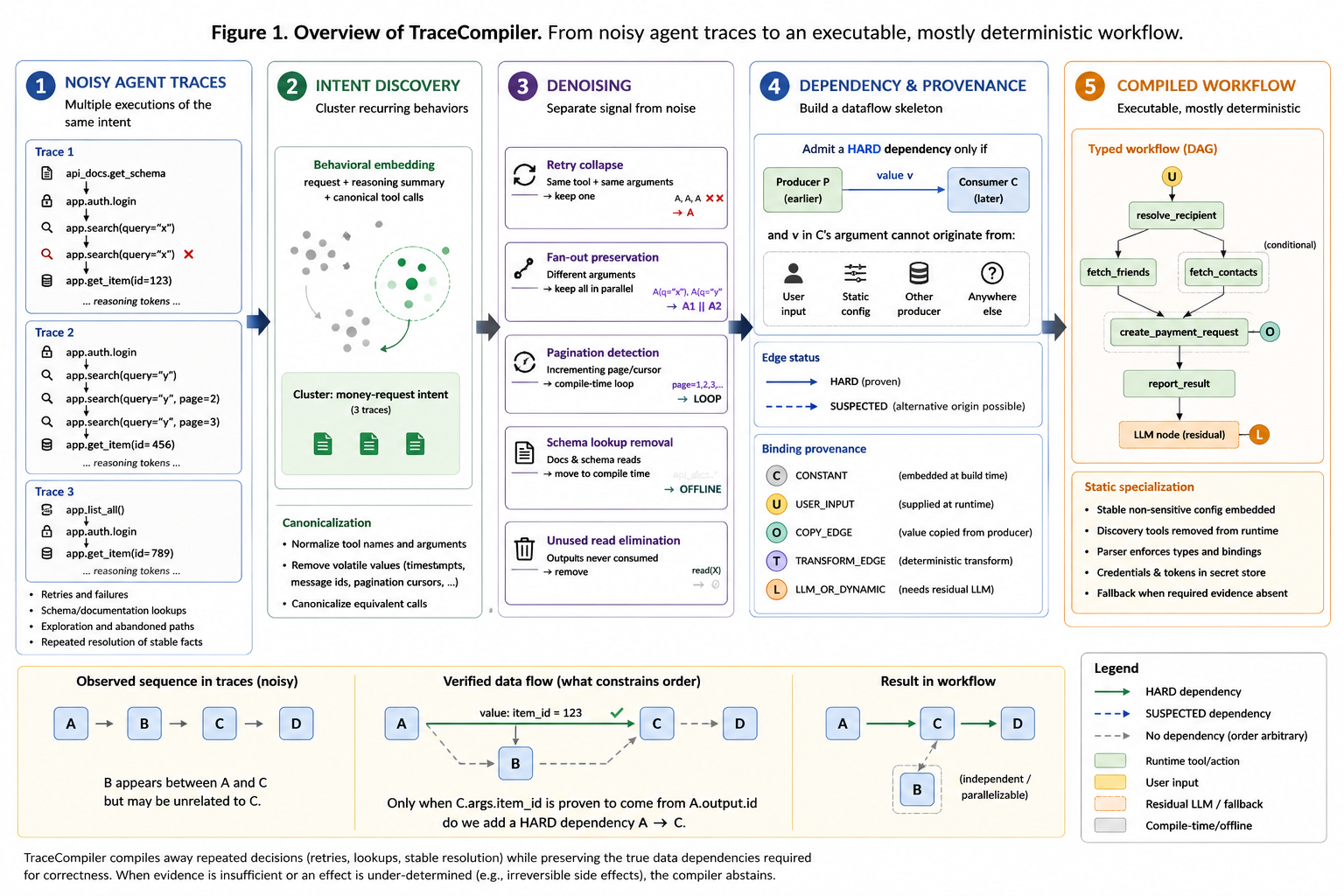}
\caption{Overview. Multiple noisy executions of one intent (1) are grouped by
behavioral clustering (2); denoising separates signal from accidental history ---
collapsing retries, preserving genuine fan-out, detecting pagination, moving
schema lookups to compile time, dropping unused reads (3); dependency
verification admits a \textsc{hard} edge $P\rightarrow C$ only when a value in
$C$'s argument provably originates from $P$ and no alternative source (4); the
result is a typed workflow in which evidence-backed decisions are compiled away
and only genuinely open choices remain as LLM nodes (5). The bottom band states
the central principle: observed adjacency does not imply dependency. Tool names
are illustrative.}
\label{fig:overview}
\end{figure*}

An exploratory trace contains both reusable procedure and accidental execution
history. The reusable part includes the necessary tools, true data dependencies,
branch conditions, stable configuration, and value transformations. The
accidental part includes retries, abandoned exploration, stylistic ordering,
schema-discovery lookups, and repeated resolution of facts stable across
requests. Replaying preserves both; summarizing as textual guidance still asks an
LLM to reinterpret the procedure at runtime. We ask instead whether repeated
traces can be compiled into an executable workflow in which decisions supported
by evidence leave the model's runtime responsibility
(Figure~\ref{fig:overview}).

This differs from conventional process discovery. Event order alone is
insufficient: two calls may appear consecutively because the second consumes an
identifier the first produced, because the agent issued independent operations
sequentially, or because the first was an unsuccessful attempt. Agent logs also
expose structured arguments whose provenance determines whether an ordering is
real, and observability is often incomplete --- exported traces may record
arguments but not outputs, forcing dependencies to be inferred conservatively
from the consumer side.

TraceCompiler discovers recurring intents from a behavior-enriched
representation, separates required behavior from retries and exploration, and
applies an argument-level dependency rule: an edge $a \rightarrow b$ is retained
only when an argument of $b$ consumes a value attributable uniquely to $a$. Its
first implementation is a versioned LLM \emph{skill} --- an instruction
package a capable agent loads and follows --- rather than a trained model; it was
frozen after five adversarial audit passes and required no decision-rule changes
on corpora it had never seen.

\paragraph{Contributions.}
(1)~We formulate recurring agent-trace reuse as a compilation problem over
workflow structure, parameter provenance, and residual LLM decisions, under
partial observability. (2)~We present a pipeline for unsupervised intent
discovery, behavioral denoising, dependency verification, provenance
classification, and executable workflow generation. (3)~We formalize and
evaluate a conservative argument-level dependency rule with non-adjacent producer
scanning, explicit evidence tuples, and abstention through the \emph{suspected}
relation; producer--consumer inference by value and evidence-graded edges both
have precedents~\cite{restler,grade}, and the piece we claim is admission by
\emph{exclusion} of alternative origins with explicit abstention when exclusion
fails. (4)~We evaluate on open corpora against two ground
truths that share no machinery with each other, comparing against order-based and
recurrence-based baselines on identical data while stating what each construction
cannot test. (5)~We present compilation case studies including a refusal to
compile, execute a compiled workflow on withheld instances, and report two
corrections our own execution harness forced on us.

We also \emph{specify} scope-bound static-context injection --- a partial
evaluation of the workflow, resolving stable values at build time and
forbidding their discovery tools at runtime (Section~\ref{sec:method-provenance})
--- but do not list it as a contribution, because no result here exercises it:
neither case study reports an injected value. It is a design element awaiting
evidence, and the same applies to the \textbf{HUMAN} node type, which the
compiler is specified to emit and never does.

\section{Problem Definition}
\label{sec:problem}

A trace is a tuple $\tau = (c, u, e_1 \ldots e_n, y, z)$ with system context,
user request, execution events, final output, and metadata. An event may be a
message; an optional reasoning summary; a tool call, comprising a name and
structured JSON arguments; a tool output or exception; a state update; or a human
approval.

\paragraph{Partial observability.}
We explicitly allow traces in which arguments are recorded but tool
\emph{outputs} are not --- common in deployed logging and released benchmark
trajectories. The consequence is direct: dependencies cannot be established by
matching a producer's output against a consumer's input, and must be inferred
from the consumer side by asking whether a value could only have originated from
an earlier call. Reasoning summaries are auxiliary evidence, never ground truth.

\paragraph{Intent clusters.}
The unit of compilation is an \emph{intent cluster} $C = \{\tau_1 \ldots
\tau_m\}$: traces that are variants of one task, discovered unsupervised
(Section~\ref{sec:method-clustering}). A single trajectory conflates required
structure with one agent's search behavior, and cross-trace induction outperforms
per-trajectory induction~\cite{wiseflow,skilldisco}. Traces are noisy in
characteristic ways: agents retry ill-parameterized calls, read documentation
before acting, perform lookups whose outputs are never consumed, interleave
exploration with essential work, and occasionally mix unrelated intents in one
conversation.

\paragraph{Output workflow.}
The compiler emits $W = (V, E, B, G, I, O)$: typed nodes, control-flow edges,
parameter bindings, guards, and schemas. Node types are START/END, TOOL,
TRANSFORM, DECISION, LLM, and HUMAN. Every argument binds to a literal, a
workflow input, a prior output, a deterministic transform, or an LLM result; the
binding's provenance class (Section~\ref{sec:method-provenance}) determines
whether runtime resolution needs a model. A HUMAN node is emitted in exactly one
circumstance: when a node with an irreversible external effect has a binding the
analysis could not resolve to a provenance class, so proceeding would require
guessing an unretractable value. Neither case study emits one, and we flag that
rather than let the type list imply coverage. A definition is compiled per intent
but \emph{instantiated} per user: bindings stable for a user (identity, team,
configuration, credentials) resolve once at build time.

\paragraph{Desired properties, and which we measure.}
A useful compiled workflow should satisfy \textbf{fitness} (it explains the
successful traces), \textbf{precision} (it permits no unsupported behavior),
\textbf{generalization}, \textbf{minimality}, \textbf{determinism},
\textbf{robustness}, and \textbf{auditability}. The first four names are borrowed
from process mining's evaluation quartet~\cite{vanderaalst2016}, and we do not
measure fitness and precision with that field's instruments --- our artifacts are typed programs
with guards and transforms rather than Petri nets, so replay measures would need
a translation whose faithfulness itself needs defending. What we quantify is
minimality (observed versus runtime calls), determinism (residual LLM nodes),
generalization (execution on withheld instances,
Section~\ref{sec:results-execution}), and auditability, structurally. We make no
conformance claim.

The compiler's design priorities trade runtime LLM nodes, tool-call cost,
structural complexity, and replay error against fitness and validation
thresholds. We state this as a design orientation rather than an objective
function: no weights are fitted, no thresholds assigned, and no number in this
paper is the value of such an expression. Its role is only to fix which
quantities are costs. The premise that a capable model may spend effort once,
amortized over repeated executions, is likewise a design premise, not a measured
result --- we do not report offline compilation cost.

\section{Method}
\label{sec:method}

\subsection{Intent discovery}
\label{sec:method-clustering}

Each conversation is embedded as one document combining the initial request
(weighted double as the intent carrier) with the agent's round-by-round
execution, each round's reasoning interleaved with its canonicalized tool calls;
interleaving preserves procedural order rather than treating tools as a bag.
Threshold agglomeration (cosine, average linkage) gives interpretable control
over conservative under-merging, its single parameter reading as a minimum
intra-cluster similarity. Under-merging is less harmful than over-merging: two
clusters of one intent compile to two valid workflows, whereas over-merging
contaminates extraction. Clusters below a support floor are set aside as ``not
yet compilable'' rather than forced into neighbors. An idempotent pre-pass
detects conversations mixing intents (coarse embedding drift, then mandatory
content reading) and splits them; borderline cases are left unsplit.

One applicability condition surfaced, and it applies to T1 itself: on corpora
whose assistant turns are scripted boilerplate, the interleaved text becomes
shared noise that collapses clusters, so the embedding falls back to
request-plus-tools. This is a stated condition selected by a corpus property
observable before clustering, not a hyperparameter fitted to the outcome.

\subsection{Normalization and behavioral denoising}
\label{sec:method-denoising}

Raw traces name the same logical operation inconsistently, so tool identifiers
are mapped to a canonical \texttt{app.action} vocabulary --- extending to
argument paths --- before any comparison. This has a cost we quantify later:
collapsing distinct read and write operations on one resource can erase the very
argument that proves a dependency.

Each action is labeled \textsc{essential}, \textsc{recovery},
\textsc{exploratory}, \textsc{redundant}, \textsc{dead\_output},
\textsc{policy\_required}, or \textsc{unknown} (kept, never silently dropped). An
action is \textsc{dead\_output} when the observable trace contains no downstream
argument, retained decision, or final-answer element attributable to its output;
schema-discovery lookups are the canonical example, informing the agent's next
call at run time but binding no downstream runtime value.

The costliest confusion is between repetition-as-noise and
repetition-as-structure, and call counts cannot separate them. The discriminator
is \emph{argument comparison}. Repetitions whose arguments jitter toward a final
successful call are retry noise, subsumed by the last occurrence; repetitions
whose arguments are systematically distinct (one search per song) are genuine
fan-out and compile to parallel invocations. A literal counter advancing across
repeats (\texttt{page\_index} 0,1,2 with constant query) is a loop, not
independent calls and not a data dependency. The stage does not select the
shortest trace and does not delete on suspicion: removal requires positive
evidence.

\subsection{Argument-level dependency verification}
\label{sec:method-dependencies}

Observed adjacency is never evidence of dependency; only proven data flow
constrains the compiled DAG. For calls $a \prec b$ we admit $a \rightarrow b$
only if some argument of $b$ \emph{consumes} a value only $a$ could have
produced. Let $v$ be bound to an argument path of $b$; the edge requires (i)~$v$
traceable to $a$'s output, and (ii)~every alternative origin of $v$ excluded ---
the user's utterances, embedded static context, tool-schema defaults, and the
outputs of every other call preceding $b$. Condition~(ii) separates verification
from pattern matching.

Every retained hard edge carries a tuple $\langle$\textit{consumer},
\textit{arg\_path}, \textit{value}, \textit{presumed producer},
\textit{exclusions}$\rangle$, so a reviewer can trace any constraint to concrete
values. When the evidence admits a plausible alternative --- most commonly two
type-compatible producers preceding one consumer, or a value that could have come
from an unrecorded request --- the edge is downgraded to \emph{suspected}:
reported, imposing no ordering constraint. The producer scan covers \emph{all}
earlier calls: credentials resolved once are consumed after unrelated intervening
calls, and restricting the scan to adjacent pairs would both miss such edges and
inflate false ones. Each relation receives one of \textbf{hard}, \textbf{conditional
hard}, \textbf{suspected}, \textbf{independent/parallel}, or \textbf{arbitrary
order}; only the first two constrain the DAG.

Two evidence regimes arise. When concrete values are recorded, consumption is
established by value matching. When values are opaque, the method falls back on
co-reference of the opaque value, temporal precedence, and elimination of
type-compatible alternatives --- a strictly weaker regime in which the suspected
downgrade carries the safety burden. Our evaluation exercises both.

\subsection{Provenance and static context}
\label{sec:method-provenance}

Every binding receives one of \textsc{constant} (stable across the intent's
traces, not attributable to request text), \textsc{user\_input},
\textsc{copy\_edge} (the consumer side of a proven hard edge),
\textsc{transform\_edge}, or \textsc{llm\_or\_dynamic} (genuinely semantic
bindings, which alone justify a runtime LLM node). Classification is behavioral,
not name-based. Authentication needs care: account credentials are scope-stable
and may resolve once, but the \emph{bearer token} an auth call returns is a
session artifact that can expire, so it is a \textsc{copy\_edge}, never a
\textsc{constant}. We therefore separate a \emph{cold} path, where a session is
initialized, from a \emph{warm} path where the token is already held; reporting
reductions against the warm path avoids overstating amortization a fresh session
would not enjoy.

\textsc{constant} bindings admit a stronger optimization than literal
substitution: at build time the compiler executes each discovery tool once,
embeds resolved values into the parser node's prompt with source and usage
annotations, and forbids those tools at runtime, escalating rather than silently
rediscovering if an embedded value is rejected. Injection is an outcome of
analysis, not a template --- an intent whose every argument is
\textsc{user\_input} is a legitimate outcome, and forcing injection where nothing
is stable would be the overfitting this stage exists to avoid.

\subsection{Compilation, validation, and runtime}
\label{sec:method-compilation}

The analysis is compiled into an executable definition, then treated as a
candidate program rather than trusted output. Generation is driven by a
structured specification stating \emph{what} was proven, never the builder's
syntax. Output is validated mechanically (schema conformance, topic wiring,
variable definition before use) and structurally against the specification;
failures become diagnostics under a bounded repair loop, and a candidate that
cannot satisfy its specification is rejected. When an irreversible side effect is
under-determined by the traces, the compiler emits no executable graph and
escalates.

The runtime router uses a request-only representation. A match above a
conservative similarity threshold dispatches to the compiled workflow; unmatched
requests fall back to the general agent. The asymmetry is deliberate: a false
negative costs one agentic execution, the status quo, whereas a false positive may
execute the wrong procedure. Routing accuracy is not evaluated here, which
matters because the router gates a workflow whose terminal step moves money.

\section{The Compiler Skill}
\label{sec:skill}

The compiler's v1 engine is a \emph{skill}: a versioned instruction package a
capable LLM agent loads and follows over a cluster's traces. The analysis stages
involve judgments (retry vs.\ variation, evidence sufficiency, injection
candidacy) costly to encode exhaustively in symbolic form but specifiable as
procedure, testable adversarially, and revisable quickly. The method is
engine-agnostic, and every execution logs the (cluster, workflow, evidence)
triple a learned compiler could train on. The package carries an explicit input
contract (stop and report on schema mismatch, rather than adapt silently), and
identifies its target cluster by describing the intent, execution count,
dominant tool families and variants rather than by a numeric id, which is
unstable across re-clusterings. It comprises ten steps (0--9), condensed in
Appendix~\ref{app:skill}; the full package is released with the artifact.

We should be precise about what ``frozen'' means. The released package is
versioned (v1.1.0, with a changelog), but the v1.0.0 text that produced every
result here carried no version field and no audit log. The claim that it was
frozen after five adversarial audit passes is therefore \emph{our testimony, not
a checkable property of that release}, and one post-freeze commit did touch a
helper script's import path, though no decision rule. A reader should treat the
freeze as an assertion; versioning begins with v1.1.0 so the ambiguity does not
recur. The skill's taxonomies are also finer than this paper's prose, in ways
Appendix~\ref{app:skill} sets out.

\section{Experimental Setup}
\label{sec:setup}

\paragraph{Data.}
\emph{T1}~\cite{t1}: template-generated task dialogues across nine travel domains
whose reference plans are executable code. T1 comprises 13{,}500 dialogues from
60 templates per domain family, split 15 train / 5 validation / 40 test
templates. \textbf{We use the training split only}: our adapter converts its
3{,}375 dialogues (14{,}250 plans, zero parse failures) and extracts a def--use
dependency graph. Where we say ``the corpus'' for T1 we mean that split --- 25\%
of the dialogues and the quarter of the template inventory with least structural
diversity; the 40-template test split is untouched, and running the deterministic
evaluation on it is the obvious next strengthening. def--use yields 22{,}850 edge
occurrences, deduplicating to \textbf{15{,}775} distinct (conversation, producer,
consumer) edges over the \textbf{3{,}350} dialogues containing at least one
dependency. Every T1 figure is computed against the deduplicated 15{,}775.

\emph{AppWorld}~\cite{appworld}: an executable benchmark of nine applications and
457 APIs, with released trajectories over 56 recurring scenarios, each
instantiated three times. Trajectories record arguments but not outputs, placing
them in the value-opaque regime.

\emph{Hermes-Function-Calling-v1}~\cite{hermesfc}: a single-turn function-calling
corpus (configuration \texttt{func\_calling\_singleturn}) stressing
argument-schema normalization. It should not be confused with
ToolMind~\cite{toolmind}, which is \emph{multi-turn} and builds a function graph
from parameter correlations --- close to the opposite of the single-turn corpus
our negative result concerns, and, as we note under Transfer
(Section~\ref{sec:discussion}), well matched to the transfer experiment this
paper lacks.

\paragraph{What is model-dependent.}
The blind protocol of Section~\ref{sec:results-e4} and the two case studies were
executed by a frontier LLM agent whose version we did not record; an LLM-driven
compiler is not deterministic across runs or versions, so those results are one
observation of a stochastic procedure, not reproducible constants. Everything
else is model-free: masking, the mechanized rule and its baselines,
replay-established references and their scoring, discovery metrics, and the
end-to-end execution are deterministic released commands. This is the main reason
we added the mechanized rule --- it places the central accuracy claim on a
footing independent of model version, at the cost of a cruder alternative-origin
test.

\paragraph{Discovery parameters.}
Conversations are embedded with a general-purpose sentence encoder
(\texttt{all-mpnet-base-v2}); agglomeration uses cosine distance and average
linkage at minimum intra-cluster similarity \textbf{0.45}; the support floor is
\textbf{5}. The evaluated sample is a balanced subsample of \textbf{80 dialogues
per domain} (720 of 3{,}375), drawn before clustering. We did not tune threshold
or floor against the labels, but neither did we hold out a set for choosing them,
so these are defaults rather than tuned optima; no sensitivity sweep is reported.

\paragraph{Uncertainty.}
We attach Wilson 95\% intervals to every precision and recall figure, since those
estimates are intended to generalize and the normal approximation is unreliable
near 1. We do not attach intervals to descriptive proportions of a fixed artifact
(schema-discovery share, replay success, purity, coverage): these are censuses of
released data, not samples, and an interval would imply a sampling model that does
not exist. ARI and NMI are reported without intervals for the same reason. Where
a distribution is the honest summary we give median and interquartile range. One
caveat we do not resolve: T1's edges are nested in templates, so treating them as
independent Bernoulli trials makes our intervals narrower than a
template-clustered resampling would.

\paragraph{Research questions.}
\textbf{RQ1}: does the skill compile recurring trajectories into correct
specifications, recovering true dependencies and refusing unsafe ones?
(Section~\ref{sec:results-case}) \textbf{RQ2}: does argument-level verification
recover true dependencies against two reference constructions that share no
machinery with each other? (Sections~\ref{sec:results-e4},
\ref{sec:results-observable}) \textbf{RQ3}: does
compiled structure hold outside the executions it came from
(Section~\ref{sec:results-execution}), and do the stages transfer to a foreign
corpus (Transfer, Section~\ref{sec:discussion})? The rule is compared against
adjacency, all-pairs, and a recurrence measure on identical data. We do not run
competing systems' full pipelines, which target different outputs and
environments; the comparison isolates structural criteria.

\section{Results}
\label{sec:results}

\subsection{Compiling open AppWorld intents}
\label{sec:results-case}

We compile two recurring intents chosen for structural contrast: a chain with an
alternative branch, and a fan-out coordinating two applications. Both are drawn
from released ReAct-agent trajectories, each intent recurring across three
instances. Because AppWorld records arguments but not outputs and masks bearer
tokens at source, both are compiled in the value-opaque regime, and we flag every
dependency whose provenance the masking leaves ambiguous.

\begin{figure}[t]
\centering
\includegraphics[width=\linewidth]{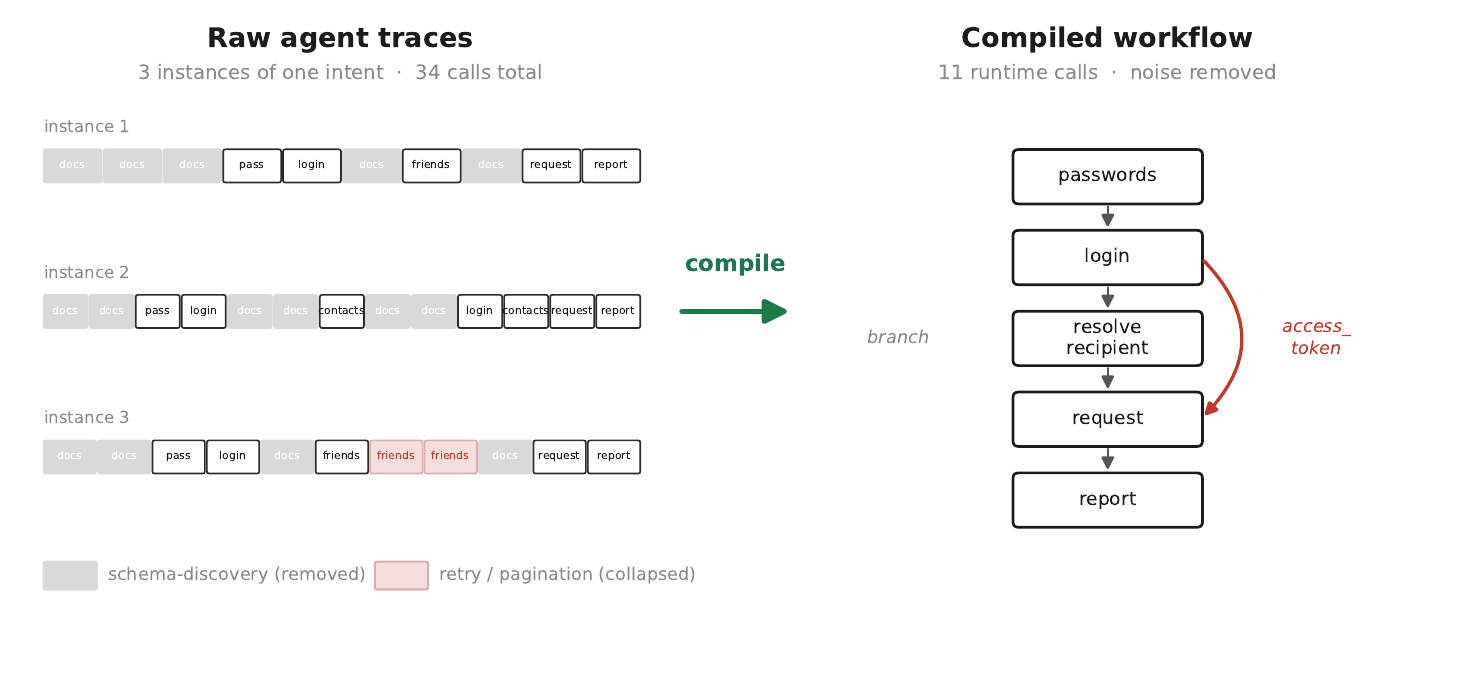}
\caption{Compiling the Venmo money-request intent (AppWorld 024c982): raw traces
(left) versus compiled workflow (right). Grey cells are schema-discovery
lookups, red are retries and pagination, white are essential steps; roughly half
of this cluster's calls are removable noise. The \texttt{access\_token} dependency (red) is
recovered as a hard edge and recipient resolution left as a branch
(\textsc{llm\_or\_dynamic}). The three instances' 34 calls compile to 11 runtime
calls (3, 3, 5).}
\label{fig:venmo}
\end{figure}

\begin{figure}[t]
\centering
\includegraphics[width=\linewidth]{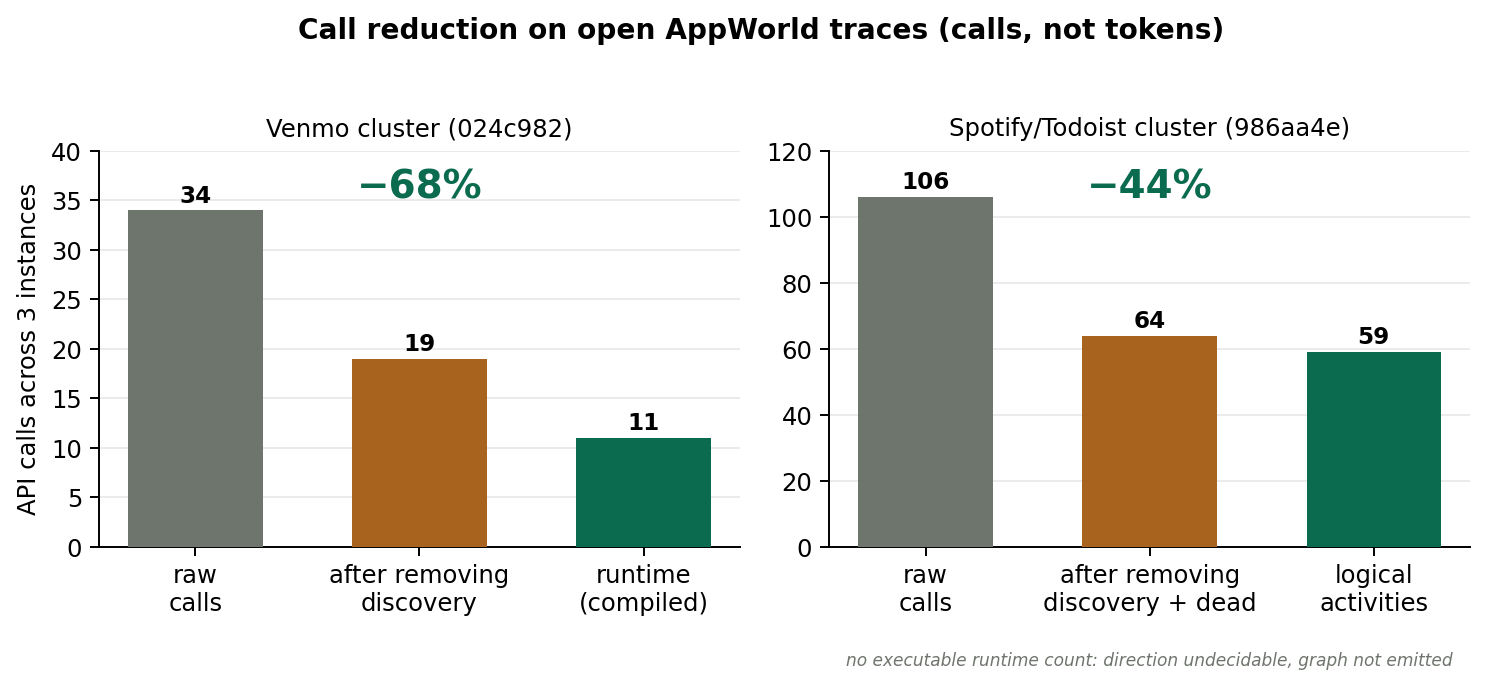}
\caption{Call reduction on open AppWorld traces (calls, not tokens). Venmo
compiles to a runtime path ($34 \rightarrow 11$ across three instances).
Spotify/Todoist reduces to logical activities; no executable runtime count is
reported because the mutation direction is under-determined.}
\label{fig:reduction}
\end{figure}

\paragraph{Intent 1: request money on Venmo (Figure~\ref{fig:venmo}).}
The three instances share a skeleton --- resolve the requesting account,
authenticate, resolve the recipient, create the request, report --- and differ
where the method predicts. Across them the agent issues 34 API calls, of which
roughly half are schema-discovery lookups that bind no runtime value and are
removed.\footnote{We do not state an exact schema-discovery count here. Our
released activity tables for this cluster are incomplete --- they enumerate 30
of the 34 calls --- and the counts recoverable from them (11, or 13 by the
spec's own summary line) disagree with each other. The reduction to 11 runtime
calls is unaffected, being fixed by the compiled path rather than by the
discovery count, but we prefer an approximate figure we can support to a precise
one we cannot. Completing those tables from the raw trajectory is a pending
correction, flagged in the released artifact.} One instance
retries an authentication with a corrected credential (collapsed to one logical
auth), and one pages through the friend list three times (\texttt{page\_index}
0,1,2 with constant query --- one pagination loop, not three independent calls).
Under a warm authenticated session the three instances require 3, 3, and 5
runtime calls: $34 \rightarrow 11$ (Figure~\ref{fig:reduction}). The third
instance's 5 counts the pagination loop's three iterations as three
\emph{issued} calls, since the runtime still pays for each page; as
\emph{activities} it is one node, which is why the workflow specification
records nine. We count issued calls throughout so the reduction is not
flattered by folding a loop into a single unit.

The analysis recovers the token flow as a hard edge: \texttt{venmo.auth} produces
an access token consumed by \texttt{venmo.payment\_requests}, absent from every
prior call's user-supplied arguments and producible by no other tool. A
non-adjacent credential edge is also recovered --- the phone-branch password read
from the same multi-account bundle nine calls earlier, where an adjacency-only
reading would have attributed it to an intervening documentation call. Recipient
resolution is the cluster's branch: two instances resolve through
\texttt{venmo.search\_friends} (\textbf{Branch~A}), the third through
\texttt{phone.search\_contacts} (\textbf{Branch~B}); we use those labels
throughout.
Because the traces never show which condition selects the branch, the compiler
marks the selection \textsc{llm\_or\_dynamic} rather than baking in a Venmo-first
policy --- a deliberate abstention, since for irreversible money movement an
automatic two-address-book search on a bare first name would be unsafe. The
recipient-email edges are reported \emph{suspected}: the user request is absent
from the export, so ``the email could not have come from the user'' is plausible
inference, not formal exclusion.

\begin{figure}[t]
\centering
\includegraphics[width=\linewidth]{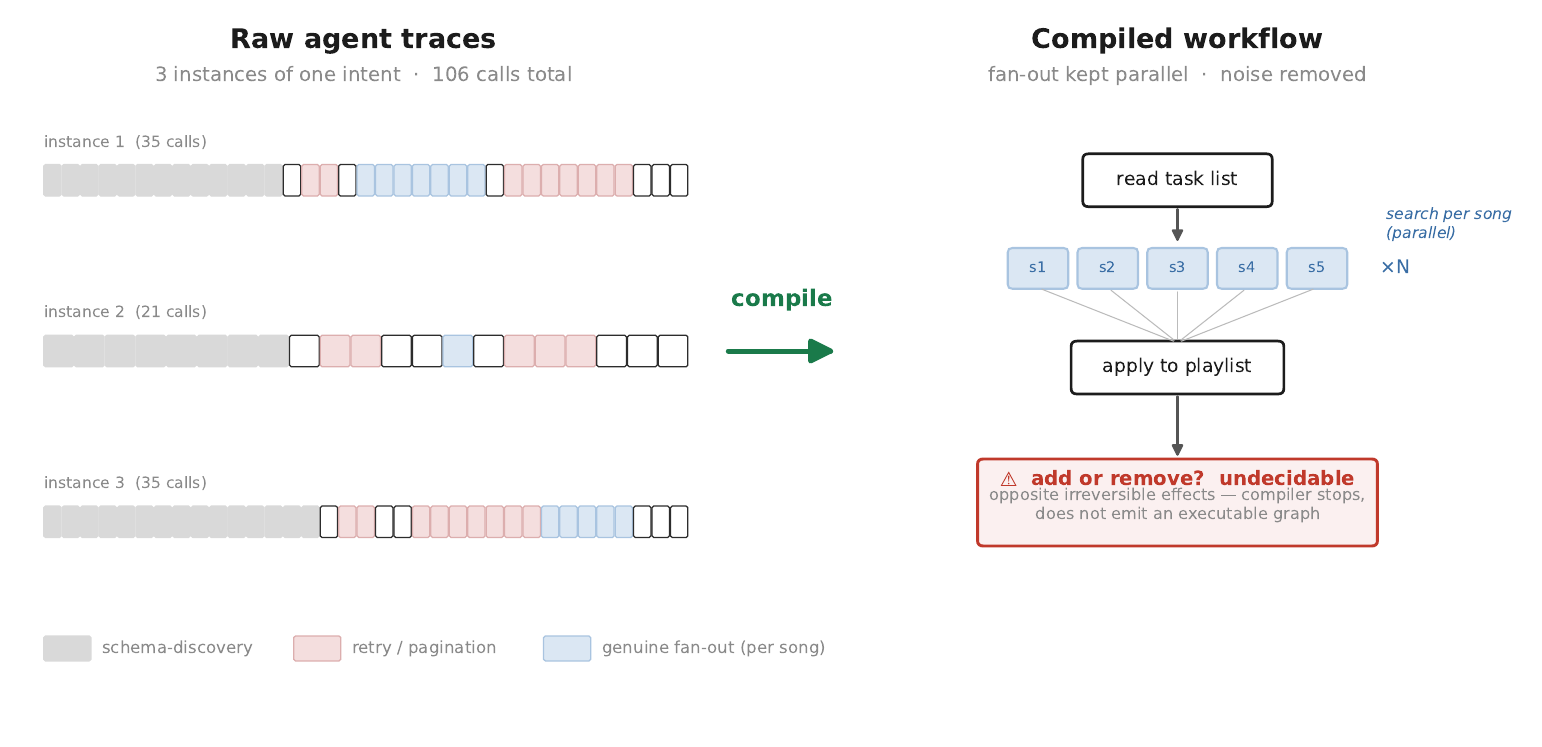}
\caption{Compiling the Spotify/Todoist intent (AppWorld 986aa4e): raw traces
(left, 106 calls) versus compiled workflow (right). Grey is schema-discovery, red
retry/pagination, blue genuine per-song fan-out, white essential. The workflow
keeps per-song searches parallel ($\times N$) but reaches a blocking gap: whether
each song is added to or removed from the playlist is under-determined. Because
these are opposite irreversible effects, the compiler emits no executable graph.}
\label{fig:spotify}
\end{figure}

\paragraph{Intent 2: apply a to-do list's songs to a playlist
(Figure~\ref{fig:spotify}).}
The second intent coordinates two applications and is dominated by repetition,
stress-testing the retry-versus-variation rule. The same rule yields opposite
verdicts on tools with nearly identical repeat counts: \texttt{spotify.songs},
fired up to seven times per instance with a distinct \texttt{query} each time, is
genuine fan-out (fully parallel); \texttt{todoist.projects}, fired twice with
byte-identical arguments and never consumed, is a dead read and is dropped; a
credential re-fetch and an authentication repair are collapsed as retries.

Across three instances the intent issues 106 calls, and the collapse rule matters
enough to give both readings. Removing the 36 \texttt{api\_docs} lookups (34\%)
leaves 70; Figure~\ref{fig:reduction} additionally removes 6 dead reads and then
5 retry/pagination repeats, reaching \textbf{59} logical activities. A stricter
reading of our own retry rule collapses \emph{every} byte-identical repeat within
an instance --- such a call being by construction either a retry subsumed by its
last occurrence or a dead re-read --- which removes 25 rather than 11 and leaves
\textbf{45}. The 14-call difference is exactly the repeats the conservative ladder
retained. We report both, since the figure is meaningless without the rule that
produced it. One caveat on the stricter reading: it collapses byte-identical
calls regardless of position, so a legitimate re-read after an intervening
mutation would be counted as noise.

This intent exposes two honest limits. First, \emph{canonicalization cost}: the
export collapsed \texttt{show\_playlist}, \texttt{add\_song\_to\_playlist}, and
\texttt{remove\_song\_from\_playlist} into one canonical action and dropped the
playlist and song identifiers, so the load-bearing edge --- resolved song feeding
the playlist mutation --- can only be reported \emph{suspected}: sound by API
semantics but unconfirmable from masked arguments. Second, the mutation's
\emph{direction} --- added to or removed from the playlist --- is
under-determined: all three instances read both the add and remove documentation,
and no call records a direction. Because these are opposite, irreversible changes
to a user's library, the compiler emits \emph{no executable graph} and reports the
ambiguity. Declining to compile an unsafe under-specified side effect is
robustness, not failure.

\subsection{How much of a trace is removable, corpus-wide}
\label{sec:results-noise}

Two intents invite the objection that they were chosen because they reduce well.
We therefore measure the two evidence-determined denoising categories ---
schema-discovery calls and byte-identical repeats within an instance --- across
\emph{all} 56 recurring scenarios (14{,}128 calls). Genuine fan-out is excluded
by construction, since a per-item call carries a distinct argument each time.

Per scenario the median removable fraction is \textbf{51.4\%} (IQR 41.9--74.1\%,
range 12.1--99.3\%); 51 of 56 scenarios exceed 30\% and 33 exceed 50\%. We
summarize by distribution rather than corpus total because a few scenarios repeat
one read hundreds of times and would dominate the aggregate (pooled: 10.7\%
schema discovery, 71.4\% repeats). Two caveats bound this. It misses dead reads
whose arguments differ, jittering retries, and pagination loops, so it
under-counts in that direction; and it counts byte-identical repeats regardless of
intervening state changes, so it over-counts read-after-write verification.
Removability is also not a realized saving: these are calls a compiled workflow
need not issue, not a measurement of end-to-end cost.

\paragraph{Determinism, measured.}
For the one intent compiled to an executable graph, tool calls fall from 34 to 11
and the residual LLM node count is 1 --- the recipient-resolution branch the
compiler declined to resolve. In the execution experiment below that node was
replaced by deterministic rules, so 0 model invocations were needed at runtime;
that substitution is scaffolding, not compilation, and the open decision remains
in the compiled specification.

\subsection{End-to-end execution on withheld instances}
\label{sec:results-execution}

Executing on the same three instances the workflow was compiled from would
establish only that the compiler did not corrupt what it read. We therefore
report \emph{leave-one-out}: for each instance, the workflow is permitted only
the recipient-resolution branches the \emph{other two} exhibited, then executed on
the held-out instance. Where no permitted branch resolves the recipient, it must
escalate and emit no payment request.

Under this protocol the workflow passes \textbf{15 of 21} state tests. Two folds
pass completely: holding out instance~1 or~3, the training pair exhibits both
paths and the held-out instance reaches $7/7$. The third fails in the intended
direction: holding out instance~2, both training instances resolve through
Branch~A, so Branch~B is never observed, and instance~2's recipient --- not in the
Venmo friend list --- cannot be resolved by any permitted path; the workflow
escalates ($1/7$). That failing fold is what makes the protocol worth running:
structure from two executions suffices when those two exhibited the needed path
and not otherwise, which is a statement about the support a branch needs rather
than unconditional generalization.

\paragraph{A claim we withdraw, and an error in how we withdrew it.}
An earlier version asserted the alternative branch was \emph{empirically
necessary}. Single-branch execution shows otherwise: Branch~B alone suffices for
\emph{all three} instances, so a single-path workflow built on it passes every
instance, and the two-branch structure is \emph{under-determined} rather than
required. Our first correction was itself wrong, and the reason is worth recording: it reported that Branch~A suffices for instance~1 only, an artifact of
a harness that issued \texttt{venmo.search\_friends} with neither \texttt{query}
nor \texttt{page\_index}, matching against the first page of an unfiltered
listing --- discarding exactly the binding and pagination loop this section
reports the compiler recovering. With those restored, Branch~A suffices for
instances~1 \emph{and}~3 ($7/7$ each) and only instance~2 requires the phone path
(where Branch~A correctly escalates, $1/7$).

The compiler's behaviour is unchanged: marking the selection
\textsc{llm\_or\_dynamic} remains right, justified by the ambiguity rather than by
both paths being required. What the episode establishes is narrower and more
uncomfortable. Traces record which path an agent \emph{took}, not which would have
\emph{worked}, so observational data bounds claims about necessity as tightly as
claims about dependency --- our own principle, which we had applied to ordering and
not to branching. It also shows an execution harness is itself an inference
instrument: ours silently dropped a bound argument, and the resulting
``falsification'' survived a round of review before we caught it.

\paragraph{Scope.}
Argument synthesis at the residual node is performed by deterministic rules
specific to this intent's phrasing; elsewhere they mis-parse rather than erroring,
so they are scaffolding, not a contribution. This bounds what 15 of 21 measures:
it scores the compiled skeleton \emph{together with} hand-written argument
parsing, not the compiled artifact in isolation. The structural claim --- that
the recovered skeleton, branch and dependencies execute correctly against the
environment's own state tests --- is unaffected, but the number is not a
measurement of the compiler alone. The harness requires an explicit name match
and escalates otherwise, which is stricter than falling back to the first search
result: that fallback would have instantiated exactly the unsafe
bare-first-name policy this section rejects. One intent, three instances, one
application pair: an existence result about out-of-sample execution and a
demonstration that the protocol can fail, not a measurement of how often
compilation preserves correctness.

\paragraph{Why two scenarios and not fifty-six.}
This is the first question the evaluation invites and it deserves a direct
answer rather than a disclaimer. Three costs bind, and only the first is
incidental. Compiling a cluster is one long frontier-model run over every trace
in it, so 56 clusters is a materially larger spend than 2 --- but that is money,
and it does not excuse the gap. Second, and more binding: executing a compiled
workflow requires a runtime for its Workflow IR, and we have none. The Venmo
result was obtained by hand-transcribing the compiled skeleton into simulator
calls, which is feasible once and does not scale; a compile rate over 56
scenarios would be cheap, but a \emph{correctness} rate over 56 would require
the interpreter we have not built. Third, the argument-synthesis rules at
residual nodes are per-intent scaffolding, so each new scenario needs new
hand-written parsing that is not part of the method.

The consequence is a real hole and we would rather name it than let the two
case studies imply coverage. \textbf{No compile/decline rate is reported over
the 56 scenarios, and that rate --- not either case study --- is the
adoption-relevant number.} It is also the cheapest remaining experiment: the
compile side needs no runtime, only the skill and the traces, and it would turn
``the compiler declines when evidence is insufficient'' from an anecdote into a
distribution.

\subsection{Intent discovery against reference labels}
\label{sec:results-discovery}

\paragraph{Read this section knowing the table does not reproduce.}
The partition scored below was produced by a clustering run over a serialization
of T1 we no longer have. Re-running the released sweep over the shipped corpus
does not reach this operating point, and not by a small margin: the closest
measured configuration yields 12 clusters at purity 0.548, against 50 clusters
at 0.930 here, because the shipped conversations give a much tighter similarity
distribution that no threshold under either embedding variant recovers. What
survives is narrower than the numbers suggest. The \emph{metrics} recompute
exactly from the released labelling, and the qualitative failure mode is stable
across both runs --- residual merges join composite domains sharing a service
prefix, exactly as predicted --- but agreement between the two partitions is
only approximate (ARI $\approx 0.85$ at low thresholds).
Table~\ref{tab:discovery} is therefore \textbf{a measurement over a released
artifact, not a reproducible result}, and discovery is the least-supported
component of this paper. Regenerating the labelling from the shipped corpus, or
releasing the serialization that produced it, is a prerequisite for treating
these numbers as evidence. We report them because the structure they describe is
what the rest of the pipeline consumes, not because they settle anything.

Reporting purity alone would flatter an over-segmenting partition, since purity
rises monotonically as clusters shrink, so we report the full set of agreement
measures (Table~\ref{tab:discovery}). Purity 0.930 with 42 of 50 clusters entirely
label-pure says clusters rarely mix domains; ARI 0.235 says the partition does not
agree with the nine-way labelling \emph{as a partition}, because ARI penalizes
splitting a reference class; NMI 0.668 says the clusters nonetheless carry most of
the label information. A method recovering atomic domains and then subdividing
them into template-level tasks produces exactly this signature, and inspection
confirms it: the only residual merges join composite domains sharing a service
prefix, whose dialogues genuinely coincide over their first turns.

We therefore make the weaker claim the numbers support. Discovery is behaviorally
consistent with the reference labels and deliberately finer than them; it is
\emph{not} label-identical, and ARI 0.235 measures that intended difference rather
than an error rate. Purity, ARI and NMI are computed over the 357 covered
dialogues, not all 720.

\begin{table}[t]
\centering
\small
\setlength{\tabcolsep}{5pt}
\begin{tabular}{lr}
\toprule
Measure & Value \\
\midrule
Dialogues / clusters          & 720 / 50 \\
Coverage                      & 0.496 \\
Purity (over covered)         & 0.930 \\
Label-pure clusters           & 42 / 50 \\
Adjusted Rand Index           & 0.235 \\
Normalized Mutual Information & 0.668 \\
\bottomrule
\end{tabular}
\caption{Intent discovery against T1's nine reference domain labels. High purity
with low ARI and intermediate NMI is the signature of a partition systematically
\emph{finer} than the reference labelling, not one that disagrees with it.
\textbf{This partition does not reproduce from the released corpus}; see the
opening of Section~\ref{sec:results-discovery}.}
\label{tab:discovery}
\end{table}

\subsection{Dependency accuracy against reference plans}
\label{sec:results-e4}

Because the converted plans carry symbolic variable references that would reveal
the ground truth, both evaluations here read a \emph{masked} corpus: every
code-level variable reference is replaced by an opaque token
(\texttt{<REF\_$k$>}), numbered by first appearance and stable per variable within
a conversation, so co-reference survives while producer identity is hidden.
Concrete values, including cache keys, are intact. This places both evaluations in
the value-opaque regime, with no access to outputs, original plans, or the
ground-truth edge file. Applied to the shipped corpus, masking covers 14{,}925
variables across the 3{,}350 dialogues containing any.

\paragraph{Two measurements.}
The compiler's engine is an LLM, so its output is one observation of a stochastic
procedure and cannot be run over 3{,}350 dialogues at reasonable cost. Reporting
only that would leave the central claim resting on an unpinned model and a
hand-selected sample; reporting only a mechanization would not measure the system
we built. We report both, on the same corpus and ground truth.

Run blind over five clusters chosen before scoring to cover four structural
regimes, the \emph{compiler skill} attains precision and recall of $248/250 =
0.992$, Wilson interval $[0.971, 0.998]$, on 48 dialogues --- $1.6\%$ of the
corpus's edges. The two errors form a single substitution in two instances of one
template: the inference attributed a cached value to an intervening sorting call
where the annotation shows the cache received the unsorted output. This is exactly
the configuration the \emph{suspected} downgrade exists for. Because one run of a
non-deterministic procedure produced it, we do not treat the third decimal as
meaningful. Three gaps are sharper than we previously stated. The released
prompt selects its sample by a cluster identifier absent from the released
labelling; it also selects a \emph{single} cluster, whereas the protocol we
describe --- and ran --- covered five, so the released prompt does not
reconstruct the reported sample; and the 250 per-edge predictions are not
released. \textbf{This figure is therefore a reported observation, not an
auditable measurement}, and we present it as such rather than as the paper's
primary evidence. The recomputable
measurement of the same rule is Table~\ref{tab:baselines-t1}.

\paragraph{The mechanized rule, against three baselines.}
Testing what the rule does \emph{not} infer requires every method to see identical
data, so we mechanize the consumption rule --- a deterministic, model-free program
over the masked corpus --- and score it with three reference points on all
3{,}350 conversations and 15{,}775 edges (Table~\ref{tab:baselines-t1}). The
mechanization is deliberately cruder than the skill: for a consumer argument bound
to an opaque token it excludes every earlier call carrying the same token (a call
consuming a value cannot have produced it) and attributes the token to the most
recent survivor; with no schema knowledge it cannot perform the type-compatibility
judgment the evidence discipline calls for. That gap is visible ---
0.928/0.943 against the skill's 0.992 on its subset --- and is why we report both.

\textsc{adjacency} and \textsc{all-pairs} test whether \emph{order} suffices:
adjacency reaches 0.584 precision, so nearly half its edges are incidental
ordering, and all-pairs collapses to 0.231. \textsc{recurrence} tests whether
\emph{repetition} suffices, admitting an ordered pair recurring in at least a
fraction $\theta$ of its domain group's traces, swept over $[0.1,0.9]$ at best
$F_1$. Its \emph{candidate set} matters more than $\theta$, and we report both
readings, because quoting only the worse one would understate the baseline:
filtering all-pairs inherits the all-pairs precision floor and reaches 0.399
$F_1$, while filtering \emph{adjacent} pairs reaches 0.617 precision and 0.712
$F_1$. The second is the one to quote, and it \emph{exceeds} raw adjacency's
precision. Recurrence therefore genuinely improves on order; it simply does not
reach consumption.

We also correct the attribution. Calling this the operating principle of the
workflow-memory family overstated the case: none of AWM, WISE-Flow, WorkflowGen,
SKILL-DISCO or NSI~\cite{awm,wiseflow,workflowgen,skilldisco,nsi} induces
structure by thresholding ordered-pair frequency, several being LLM-driven
abstraction, and WISE-Flow's prerequisites and NSI's variable bindings are
themselves dependency-like. What we implemented is a frequency-thresholded
directly-follows dependency measure of the Heuristics Miner
kind~\cite{heuristicsminer}. The conceptual contrast with that family stands on
its own terms --- SKILL-DISCO, for instance, does not verify argument provenance
--- but it is a claim about mechanism, not a score those systems achieved.

\begin{table}[t]
\centering
\small
\setlength{\tabcolsep}{4pt}
\begin{tabular}{lrrr}
\toprule
Method & Prec. & Rec. & $F_1$ \\
\midrule
\textsc{adjacency} (consecutive pairs)   & 0.584 & 0.908 & 0.711 \\
\textsc{all-pairs} ($A\!<\!B$)           & 0.231 & 0.998 & 0.376 \\
\textsc{recurrence}, all-pairs cand.     & 0.253 & 0.945 & 0.399 \\
\textsc{recurrence}, adjacent cand.      & 0.617 & 0.842 & 0.712 \\
\midrule
argument-level, most recent survivor     & 0.928 & 0.943 & 0.936 \\
argument-level, unique-candidate only    & 1.000 & 0.071 & 0.133 \\
\bottomrule
\end{tabular}
\caption{Dependency recovery on T1's training split: 15{,}775 deduplicated
def--use edges over 3{,}350 conversations, recurrence at $\theta{=}0.6$. Wilson
95\% precision intervals in row order: $[.578,.590]$, $[.228,.235]$,
$[.250,.257]$, $[.610,.623]$, $[.924,.932]$, $[.997,1.000]$. \textbf{The rows do
not see equivalent inputs}: order-based rows read only tool order, while
argument-level rows additionally read masked co-reference, which this corpus
supplies for free. The skill's 0.992 is deliberately not a row --- different
executor, different sample.}
\label{tab:baselines-t1}
\end{table}

\paragraph{What this corpus cannot test.}
One property bounds every argument-level number above. The masked corpus
contains 22{,}975 \emph{consumption relations} --- (consumer call, opaque
token) pairs --- which is not the 22{,}850 edge \emph{occurrences} of
Section~\ref{sec:setup}. The two counts differ by 125: 75 are multiplicities
that the edge file collapses (one variable re-consumed by the same tool later in
a dialogue), and 50 are relations the annotation does not license at variable
granularity. The completeness figure then depends on which granularity is asked
for, and both are worth stating. At the \emph{tool-pair} granularity our ground truth and every
row of Table~\ref{tab:baselines-t1} use, \textbf{all 22{,}975 relations
correspond to a true def--use edge}. At the stricter \emph{per-variable}
granularity --- does the consumer consume \emph{that} variable in the annotation
--- 22{,}925 of 22{,}975 do ($99.8\%$); the 50 exceptions are two template
families repeated across instantiations, and are annotation gaps rather than
inference errors. Either way a token's presence in an argument is an almost
perfect dependency signal, and the task measured is \emph{producer attribution},
not edge \emph{detection}: there is little opportunity to emit a false positive
except by mis-ranking candidates.

Two caveats on how much even that establishes. Every relation has at least one
candidate producer --- the count of zero-candidate relations is exactly zero ---
so the abstention branch an alternative-origin exclusion is meant to trigger
never fires on this corpus: a qualifier such as ``with at least one candidate
producer'' would describe a filter that removes nothing. And the
tool-pair test is weak on its own: a mean of 8.2 candidates survive per relation
and a randomly chosen one satisfies it 36\% of the time.

That last figure invites the objection that the metric is simply forgiving, so
we test it directly by ablating the attribution step while holding the exclusion
fixed (Table~\ref{tab:ablation}). It is not forgiving: choosing at random among
the same survivors drops precision from 0.928 to 0.403, and choosing the
earliest to 0.146. But the ablation also shows where the accuracy comes from,
and it is not where the method's rhetoric points. The exclusion filter is nearly
vacuous here; the operative decision is \emph{recency}, and the mechanized rule
is best described as ``most recent survivor'' rather than as an exclusion test.
The exclusion discipline the paper argues for is exercised by the skill, not by
the mechanization, which is the price of making the central number model-free.

\begin{table}[t]
\centering
\small
\setlength{\tabcolsep}{4pt}
\begin{tabular}{lrrr}
\toprule
Candidate-choice policy & Prec. & Rec. & $F_1$ \\
\midrule
most recent survivor (as reported) & 0.928 & 0.943 & 0.936 \\
random survivor                    & 0.403 & 0.440 & 0.420 \\
second-most-recent survivor        & 0.220 & 0.214 & 0.217 \\
earliest survivor                  & 0.146 & 0.073 & 0.097 \\
all survivors (no attribution)     & 0.365 & 1.000 & 0.535 \\
\bottomrule
\end{tabular}
\caption{Ablating attribution while holding the exclusion filter fixed. Every
row reads only ref-bearing arguments and drops candidates carrying the ref;
rows differ only in which survivor receives the edge. The collapse under random
choice shows the metric is not forgiving; the gap between rows shows recency,
not exclusion, is doing the work.}
\label{tab:ablation}
\end{table}

Two further consequences follow. The last row's precision of 1.000 is close to
analytically forced rather than empirically earned: the call defining a variable
never carries that variable in its own arguments, so it always survives the
carrier exclusion, and when one candidate survives it \emph{is} the definer. We
name that row for what it computes --- unique-candidate --- rather than
``selective'', because it emits a hard edge only when exactly one candidate
survives, which with a mean of 8.2 survivors is a degenerate criterion and
explains its recall of 0.071. It is a lower bound on abstention, not the
method's abstention behaviour, which rests on type compatibility and semantic
exclusion the mechanization cannot perform. Second, measuring \emph{detection}
would need a construction that can produce false positives --- decoy tokens, or
masking every identifier-shaped literal including user-supplied ones. We have
not built that, and it is the missing construction in this evaluation.

What the 0.928 does establish, then, is producer \emph{attribution} under
co-reference at high accuracy, on a corpus where detection is trivial and the
exclusion test is close to vacuous. That is a necessary condition for the rule,
not a demonstration of it --- and it is measured on synthetic
template-generated dialogues whose dialogues and reference plans came from one
generator, so incidental regularities are more systematic than in human or agent
traces.

\subsection{Token attribution on AppWorld}
\label{sec:results-observable}

AppWorld's logs mask bearer tokens at source, so token dependencies --- the
dominant type in these traces --- are invisible to value matching. Because the
simulator is deterministic, replaying a trajectory recovers the real return values
the log omits; we replay each call, capture outputs, and substitute real values
for masked placeholders as execution proceeds. On the resulting trace an edge is
established by exact value matching.

\paragraph{This reference is not independent of the rule.}
Earlier versions described the construction as ``independent of any inference
rule'' and counted it as one of two independent ground truths. That is not
defensible and we withdraw it. Replay must \emph{choose} a value for each masked
slot, and ours prefers a token produced by the consuming application's own
authentication call. The reference is then built by matching those injected values
back to their producers, while the rule under test attributes each token to the
most recent same-application login. Injector, reference, and rule share an
assumption. All 563 edges are login-produced token edges --- the procedure
recovered no non-token identifier edge in 168 trajectories --- so this arm bounds
\emph{session-token attribution} and says nothing about the identifier flows the
case studies foreground.

This also invalidates an inference we drew from the app-blind baselines. Because
injection prefers same-application tokens, any rule ignoring application identity
is \emph{guaranteed} to disagree with the injection, so a lower score cannot
discriminate between ``application identity carries real dependency signal'' and
``it was built into the reference.'' We withdraw the claim that the gap shows a
property of the dependency structure rather than of the injection mechanism. What
the numbers support is weaker: a self-consistency check plus an ablation
quantifying how much structure application identity carries. An intervention
oracle --- replaying each consumer with its token withheld or swapped, letting the
simulator accept or reject --- would settle it, and is the clearest next step.

\paragraph{Protocol and results.}
We replay 168 trajectories over 56 scenarios spanning seven applications. Replay
executes \textbf{44.2\%} of non-documentation calls successfully (5{,}571 of
12{,}618; per-trajectory median 70.4\%); the remainder fail for faithful reasons
(an agent issuing a call before its authentication, reproduced exactly),
contributing no edge. The rate matters: more than half of non-documentation
calls contribute nothing, so the reference covers materially less of the corpus
than the per-trajectory median alone would suggest. Exact
matching yields 563 token edges, of which the opaque rule recovers 546 with 4
false positives and 17 false negatives: precision 0.993 $[0.981, 0.997]$, recall
0.970 $[0.952, 0.981]$ (Table~\ref{tab:observable}). The false negatives
concentrate on tokens consumed before any same-application login appears, where
the rule has no producer and abstains --- cases the replay independently flags as
pre-authentication token use. The false positives are the mirror case.

\paragraph{How much cross-corpus agreement this shows.}
An earlier version read 0.993 alongside T1's 0.992 as a convergence and inferred
that accuracy is a property of argument-level verification rather than of one
corpus. That inference does not survive the comparison being made correctly: the
figures were produced by different executors on different sample sizes. The
like-for-like comparison is between the two \emph{mechanical} measurements --- T1
at 0.928 and AppWorld at 0.993 --- and those do not converge; their Wilson
intervals, $[0.924,0.932]$ and $[0.981,0.997]$, do not overlap. The difference is
explicable: AppWorld's edges are almost entirely session tokens, where the
consuming application's identity is a strong signal, whereas T1's are dominated by
cache round-trips in which several type-compatible producers precede one consumer
--- precisely where the mechanization is weakest. What the corpora jointly
establish is narrower than convergence: the rule transfers across two
constructions sharing no machinery, at accuracies differing with dependency type,
and the harder type is where the skill's judgment earns its cost.

\begin{table}[t]
\centering
\small
\setlength{\tabcolsep}{4pt}
\begin{tabular}{llrrr}
\toprule
Ground truth & Rule & Edges & Prec. & Rec. \\
\midrule
T1 plans (def--use)      & skill      &   250 & 0.992 & 0.992 \\
T1 plans (def--use)      & mechanized & 15{,}775 & 0.928 & 0.943 \\
AppWorld replay (values) & mechanized &   563 & 0.993 & 0.970 \\
\bottomrule
\end{tabular}
\caption{Dependency recovery under two ground-truth constructions sharing no
machinery with each other, with the executing rule named per row. Rows are not
interchangeable and must not be averaged, and neither construction is independent
of the data producing it: T1's dialogues and plans come from one generator, and
AppWorld's reference shares an assumption with the rule it scores.}
\label{tab:observable}
\end{table}

\begin{table}[t]
\centering
\small
\setlength{\tabcolsep}{4pt}
\begin{tabular}{lcc}
\toprule
Attribution rule & Prec. [95\% CI] & Rec. [95\% CI] \\
\midrule
first login, any app   & 0.786 [.750,.818] & 0.771 [.734,.804] \\
nearest login, any app & 0.856 [.825,.882] & 0.877 [.848,.902] \\
all prior logins       & 0.728 [.696,.759] & 1.000 [.993,1.000] \\
\textbf{same-app last login} & \textbf{0.993} [.981,.997]
                             & \textbf{0.970} [.952,.981] \\
\bottomrule
\end{tabular}
\caption{Token-attribution rules against the value-observable reference (563
edges). This is an \emph{ablation} quantifying how much of the reference's
structure application identity carries, not evidence of non-circularity: replay
prefers same-application tokens, so app-blind rules must score lower. \emph{All
prior logins} attains perfect recall by construction, so only its precision is
informative.}
\label{tab:baselines-aw}
\end{table}

\section{Discussion and Limitations}
\label{sec:discussion}

\paragraph{Economics, and what is not guaranteed.}
The economics rest on an asymmetry: a capable model may spend effort analyzing a
cluster once, provided the workflow amortizes it over repeated executions. We
measure call reduction but not offline compilation cost, so we report no
break-even point. No semantics-preservation property is established either:
validation is schema and structural conformance plus state tests on one intent,
and differential replay at scale is future work. A further exposure we do not
address: removing schema-discovery calls and forbidding injected values'
discovery tools converts an API change from a self-healing event into a
silent-wrong-behavior one. We have no workflow versioning, invalidation, or
contract monitoring, and the value-level escape hatch covers stale values rather
than changed schemas, so artifact lifetime under drift is unmeasured --- leaving
both sides of the amortization argument unconstrained. Aggressive
canonicalization is the other cost: merging reads and writes of one resource, or
dropping identifier-shaped arguments, can erase the evidence a hard edge needs,
so we recommend canonicalizing conservatively.

\paragraph{When not to compile, and who authorizes.}
Several signals correctly keep intents out of the compiled set: insufficient
recurrence, no coherent argument structure, bindings that are all per-request
content, and --- demonstrated on AppWorld --- an irreversible effect whose
direction the traces leave under-determined. We do not report how often the
compiler declines, which is the adoption-relevant rate. The authorization
boundary also deserves sharper statement: on the Venmo intent the recipient
selection resolves to \textsc{llm\_or\_dynamic}, so no HUMAN node is due and a
model decides the recipient of a money request at runtime; ``escalate'' meanwhile
denotes three different things here (fall back to the general agent, stop at
compile time, emit no request). Deployments taking irreversible actions warrant
approval gating beyond what we evaluate, and BPM's answer of compensating
transactions is unmodelled because these APIs admit no reversal.

\phantomsection
\paragraph{Transfer, and a negative result.}
\label{sec:results-generalization}
We also applied the pipeline to single-turn
Hermes-Function-Calling-v1~\cite{hermesfc}, whose per-example tool vocabularies
and varying argument schemas stress canonicalization at argument-path level. It
supplies clean demonstrations of the alternative-source condition: where one text
is passed to three analysis tools, naive value matching would chain them into
spurious dependencies, whereas the rule attributes the shared value to its true
origin --- the user's request --- and leaves the calls parallel. But this is the
weakest evidence here and establishes no generalization: exactly one recurring
compilable family survives the single-turn filter, so no accuracy figure is
reportable and none is claimed. What it establishes, by absence rather than
measurement, is structural --- inter-tool data flow concentrates in multi-turn
traces and is largely absent from single-turn function calling, so such a corpus
has little for a dependency compiler to find. Demonstrating transfer needs a
second multi-turn corpus with independent annotations, and
ToolMind~\cite{toolmind} is the right candidate: multi-turn, with a synthesis
building a function graph from parameter correlations.

\paragraph{What the mechanization does and does not stand in for.}
Making the central number model-free cost something we should state plainly.
The mechanized rule implements the carrier exclusion --- a call consuming a
value cannot have produced it --- but not the rest of the evidence discipline:
it has no schema knowledge, so it cannot test type compatibility, and it does
not reason about user utterances or static context as alternative origins.
Ablating the attribution step while holding the exclusion fixed
(Table~\ref{tab:ablation}) shows the consequence: on this corpus the exclusion
leaves a mean of 8.2 candidates standing, and the accuracy is carried by
choosing the most recent of them. So the 0.928 is evidence that argument-level
co-reference plus recency recovers T1's dependencies, not evidence that
admission-by-exclusion does. The exclusion discipline is exercised only by the
skill, whose measurement is a single unaudited run. Closing that gap --- a
mechanization rich enough to perform the exclusion test, on a corpus where
detection is not trivial --- is the most direct way to put the paper's central
claim on a recomputable footing.

\paragraph{Variance, and what our own errors show.}
An LLM-driven compiler is not deterministic across runs and versions. Three
mechanisms bound the risk: the structured specification between analysis and
generation, validation with bounded repair, and evidence tuples making every
constraint traceable. Variance in the \emph{analysis} is the strongest argument
for a learned compiler trained on the triples the system logs. Separately, two of
our own claims were falsified during this work --- branch necessity, and the first
correction of it --- both by execution rather than by re-reading traces. Where a
compiled branch guards an irreversible effect, execution-checked sufficiency
should be a compile-time obligation, and the harness performing the check is
itself an inference instrument requiring validation.

\paragraph{Reproducibility and scope.}
We separate tiers rather than claim blanket reproducibility.

\begin{itemize}
\item \emph{Deterministic.} Masking, the mechanized rule and its baselines and
  ablation, the replay reference with its scoring and coverage statistics, the
  discovery metrics, and the end-to-end execution with its leave-one-out and
  branch-sufficiency protocols are model-free released commands that recompute
  their tables.
\item \emph{Procedure but not value, and not auditable.} The blind protocol and
  the case studies ran on an unpinned LLM. The prompt and masked corpus are
  released, but the prompt's sample cannot be identified in the released
  labelling and the 250 per-edge predictions are absent.
\item \emph{Not released.} The T1 adapter producing the trace and edge files, so
  a reader cannot check how def--use was computed or how masking aligned to it
  --- the step most worth auditing.
\item \emph{Known gap.} The clustering that produced the reference labelling ran
  on a different serialization and does not re-run to the same operating point
  (Section~\ref{sec:results-discovery}).
\item \emph{Regenerable, not redistributed.} AppWorld traces, because the
  benchmark ships a canary~\cite{appworldrepo}.
\end{itemize}

On scope: the corpus-wide result is measured on synthetic template-generated
dialogues, and the AppWorld measurement on traces whose dominant dependency is a
session token. Out-of-sample execution evidence is one intent over three
instances. All AppWorld traces come from one ReAct agent and all T1 dialogues
from one generator, leaving an unexamined premise --- that observational traces
are evidence about \emph{task} structure rather than one agent's habits. No
corpus demonstrates end-to-end compilation at production volume with exact
accounting, which remains the ultimate test.

\section{Related Work}
\label{sec:related}

\paragraph{Process mining, including data- and object-aware.}
Process mining derives models from event logs, evaluated through fitness,
precision, generalization and simplicity~\cite{vanderaalst2016}; Sommers et
al.~\cite{sommers2021} recast discovery as supervised graph translation over
synthetic log--model pairs. The field is emphatically not confined to control
flow, and we should be precise rather than draw a strawman. Decision mining
recovers guards routing a case at a choice point~\cite{rozinat2006}; data-aware
alignments and balanced multi-perspective conformance extend conformance to case
attributes~\cite{deleoni2013,mannhardt2016}; object-centric discovery correlates
events through shared object identifiers carried as event
attributes~\cite{ocpn}; and data-flow anti-patterns already formalize reading an
unavailable element and producing data never read~\cite{dataflowantipatterns},
which is precisely our \textsc{dead\_output} category and the ``dead fetch''
anomalies our blind analysis surfaced. The distinction we can defend is
narrower than ``control flow versus data'': those methods discover predicates
over attributes, or correlate events by identifiers the log \emph{supplies as
first-class fields}, whereas we must \emph{infer} which JSON argument value is a
reference and to which producer it belongs, from the consumer side, when outputs
are not recorded at all. Guards constrain which path a case takes; provenance
constrains which orderings are admissible.

Nor is ``observed relation need not imply dependency'' original to us. The
sharpest prior statement is Bozorgi et al.~\cite{causalpm}, who argue that rules
mined from event logs are correlational and adjust for confounding to recover
causal effect; van der Aalst's text makes the corresponding point about
directly-follows relations. We should not, however, claim the alpha-algorithm
lineage as our precedent: classical concurrency detection infers independence
from observing \emph{both} orders across a log, which requires order
variability, whereas we infer it from argument provenance within a single
ordering. Our contribution is an evidence discipline for acting on the
distinction under partial observability --- an exclusion test over alternative
origins, and abstention when the test fails --- not the distinction itself.

\paragraph{Process mining and provenance over agent traces.}
The intersection this paper occupies is already populated. Fournier et
al.~\cite{fournier2025observability} apply process and causal discovery directly
to LLM-agent trajectories to expose behavioral variability, complemented by
static analysis separating intended from unintended variation. GRADE~\cite{grade}
models an agent run as a graph with separate execution and dependency layers and
--- closest to us --- \emph{grades} each dependency edge by how it is known
(observed, declared, inferred). AgentTrails~\cite{agenttrails} converts
trajectories into provenance graphs of actions and artifacts and aligns
recurring structure across executions via a joined quotient graph. We do not
claim the graded-evidence taxonomy or the trace-to-provenance-graph construction
as new. What we add to them is a \emph{sufficiency} criterion rather than a
provenance annotation: because our graph will be executed, an edge is admitted
only after every alternative origin is excluded, and the compiler abstains when
exclusion fails. GRADE and AgentTrails produce graphs for prediction and
sensemaking and both assume observable inputs and outputs; we operate under
output opacity and emit an executable artifact.

\paragraph{Workflow memories, reuse, and skill compilation.}
Agent Workflow Memory induces recurring workflows and retrieves them as
procedural memory~\cite{awm}; WISE-Flow aggregates interactions into
prerequisite-augmented action blocks~\cite{wiseflow}; WorkflowGen routes requests
among direct reuse, rewriting, and initialization by
similarity~\cite{workflowgen}; Memp distills trajectories into step-level and
script-like procedural memory with explicit build/retrieve/update
strategies~\cite{memp}. In each, the retrieved artifact remains context for an
LLM that still selects and instantiates actions at runtime. NSI lifts traces into
logic-grounded programs, resolving ambiguity by inventing branches or predicates
rather than abstaining~\cite{nsi}; Agentic Compilation confines the LLM to a
one-shot phase producing a deterministic blueprint~\cite{agenticcompilation};
SkillCraft studies composition of atomic tools into parameterized
skills~\cite{skillcraft}; and SKILL-DISCO distills control-flow subgraphs from
successful traces in FSM-defined environments~\cite{skilldisco}.

Closest on the systems side, PreAct~\cite{preact} compiles a successful run of a
computer-use task into a state machine replayed with no per-step LLM call,
verifying at each step that the environment matches expectation and handing back
to the agent when it does not --- the same guard-and-fall-back structure as our
router, arrived at independently. It compiles a \emph{single} trace of GUI
actions with no argument-provenance analysis and no refusal to compile; our
differentiators against it are cross-trace consolidation, argument-level def--use
with exclusion, and declining under-determined irreversible effects.

Two differences locate our work generally. Their unit of abstraction is control
flow --- which transitions recur --- whereas ours is the \emph{evidence-audited}
argument binding: which runtime values resolve without model judgment, and which
observed orders actual consumption supports. And they assume successful paths
through a known transition system or a single synthesis pass, while we operate on
noisy conversational logs under partial observability. NSI and WorkflowGen also
work at binding granularity, but without a provenance-exclusion criterion. The
methods are complementary; distilling our intermediate decisions into a trained
student model~\cite{agentdistillation} is future work.

\paragraph{Dependency DAGs from specifications, and by demonstration.}
Three lines share our object but not our source of truth.
LLMCompiler~\cite{llmcompiler} has a planner emit a DAG of tool calls with
explicit inter-task argument placeholders and executes it in parallel; the DAG
is \emph{synthesized} from a task description at plan time and never verified
against observed behaviour, whereas we \emph{recover} it from traces with
per-edge evidence. RESTler~\cite{restler} infers producer--consumer dependencies
between API requests --- B follows A because B consumes a resource id A produced
--- from an OpenAPI specification plus live probing; our consumption rule is
recognizably in that lineage, and what we add is the exclusion step (ruling out
user text, injected context, defaults, and every other prior call) and operation
on logs alone, with no specification and no ability to probe. Further back,
programming-by-demonstration systems generalize a recorded trace into a
parameterized program, inferring where to insert loops and which constants to
turn into variables~\cite{rousillon}; our pagination-is-a-loop rule is a direct
descendant, differing in being multi-trace, unsupervised, and
provenance-verified.

\section{Conclusion}
\label{sec:conclusion}

We presented TraceCompiler, a skill-guided system for mining noisy LLM-agent
traces and compiling recurring behavior into executable, mostly deterministic
workflows. Its central principle is that observed execution order should not
become workflow structure unless supported by data flow.

The mechanized rule recovered dependencies at 0.928 precision and 0.943 recall
over 15{,}775 def--use-derived edges of T1's training split, against 0.711 $F_1$
for adjacency and 0.712 for a frequency-thresholded directly-follows measure on
the same data --- though, as Section~\ref{sec:results-e4} shows by ablation,
that accuracy rests on recency among surviving candidates rather than on the
exclusion test the skill performs, so it bounds the rule's mechanizable core and
not the evidence discipline itself. The skill run blind reached 0.992 on a
250-edge subset, in a single unaudited run whose per-edge predictions we did not
release, and the same rule 0.993 precision against a replay-built reference on
AppWorld reported as a self-consistency check. The frozen skill compiled recurring intents into
specifications --- recovering a token dependency, a non-adjacent credential
dependency, and a cross-application recipient path on one intent, and correctly
declining an under-determined irreversible side effect on another --- and the
compiled workflow passed 15 of 21 state tests on withheld instances, the failing
fold escalating rather than acting.

The execution harness also falsified our claim that the branch was necessary
rather than under-determined, and then, in its first version, produced a wrong
account of why, because it dropped an argument the traces bind. Together these
make a sharper point than either alone: observational traces bound claims about
necessity as tightly as claims about dependency, and the instrument used to test a
compiled workflow is itself an inference procedure needing validation. The
findings support a division of labor --- agents should resolve genuinely novel or
semantic decisions, while repeated evidence-supported behavior should be compiled
into explicit programs. Future work will build the intervention oracle and
decoy-negative construction this evaluation lacks, measure compile and decline
rates at corpus scale, and distill the skill into a trained student model.

\section*{Ethics statement}

This work analyzes three public corpora and introduces no human-subjects data.
T1~\cite{t1} and Hermes-Function-Calling-v1~\cite{hermesfc} are template- or
model-generated; AppWorld~\cite{appworld} is a simulator with synthetic personal
data, and all side-effecting executions reported here ran against that simulator,
never a live service or real account. The compiled workflows include an
irreversible money-movement operation, and the compiler's abstention and
escalation on such operations is a deliberate safety property:
Section~\ref{sec:results-case} reports a case where it declines to compile, and
Section~\ref{sec:results-execution} reports escalation in place of acting on an
unresolved recipient. Deploying compiled workflows that take irreversible actions
on real accounts warrants approval gating beyond what this paper evaluates. The
authors declare no competing interests and no external funding supported this
work.

\bibliographystyle{plain}
\bibliography{references}

\appendix

\section{The Compiler Skill (Condensed)}
\label{app:skill}

The skill is a versioned package: a specification document, four reference
procedures, and five helper scripts. This appendix condenses its operative
content and is keyed to the artifact's own step numbering, so a reader can move
between the two. Version 1.0.0 produced every result reported here; the released
package is 1.1.0, whose changelog records exactly which previously-tacit rules
were made explicit afterwards and states that no decision rule changed.

\paragraph{Input contract.}
The skill declares the trace schema it expects: one row per event with
conversation id, sequence, role, content, and a \texttt{tool\_calls} column of
\texttt{\{tool, input\}} objects. On deviation the instruction is to stop and
report, never to adapt silently.

\paragraph{Step 0 --- Validate and de-identify.}
Check the schema, and strip identifiers that must not enter the analysis.

\paragraph{Step 1 --- Contamination pass.}
Detect conversations mixing unrelated intents (coarse embedding drift, then
mandatory content reading); split at task boundaries; leave borderline cases
unsplit. Idempotent.

\paragraph{Step 2 --- Cluster coherence.}
Identify the target cluster by \emph{content fingerprint} --- example requests,
distinguishing tool, argument pattern, expected size --- never by numeric id,
which is unstable across re-clusterings. Hard abort on mismatch. Do not compile
a cluster with fewer than three independent executions.

\paragraph{Step 3 --- Normalize and compare arguments.}
Map tool identifiers to a canonical \texttt{app.action} vocabulary, extending to
argument paths. Unknown tools map to an explicit \texttt{UNMAPPED} marker rather
than being merged into the nearest name.

\paragraph{Step 4 --- Event roles.}
Label each event across traces. Retry versus genuine variation is decided by
argument comparison only, never by call counts. Unresolved events are kept,
never silently dropped. A literal counter advancing across otherwise-identical
repeats is a pagination loop, not a set of calls and not a data edge --- but an
\emph{opaque} cursor the client could not have computed is a data edge and must
be verified as one.

\paragraph{Step 4bis --- Argument-level dependency verification.}
For each consumer argument, scan \emph{all} earlier calls as candidate
producers. A hard edge requires all four of: traceability of the value to a
producer output path; uniqueness, meaning no user-input span, constant, static
context, default, or other prior producer plausibly explains the same value;
path compatibility of types; and a counterexample check. Record every accepted
\emph{and rejected} edge as a tuple $\langle$consumer, argument path, observed
value, proposed producer, producer output path, alternative origins checked,
verdict, confidence$\rangle$. A plausible alternative downgrades to
\textsc{suspected}, which imposes no ordering constraint.

\paragraph{Step 5 --- Provenance.}
Assign every argument an origin class. Classification is behavioral, never
name-based. Session credentials need care: account credentials may be
scope-stable, but a token a call \emph{returns} is a session artifact and is a
copy edge from the authenticating call, never a constant.

\paragraph{Step 6 --- Static-context injection.}
For each eligible value: resolve once at build time, embed into the parser
node's prompt annotated with source and usage, and forbid the discovery tool at
runtime, escalating rather than silently rediscovering. Weak-evidence candidates
are flagged, not embedded; an empty injection set is a valid outcome.

\paragraph{Steps 7--9 --- Graph, IR, validation.}
Build the smallest graph the traces support: only proven hard edges constrain
order, independent nodes stay parallelizable, retries collapse into policy,
fan-out is explicit. Emit the Workflow IR, then validate it mechanically and
structurally against the specification under a bounded repair loop.

\paragraph{Stop conditions.}
The skill abstains when a cluster cannot be segmented, when fewer than three
traces support a structural claim, when schemas are missing and values opaque,
when multiple producers remain plausible, or when a static value cannot be
scoped. One condition blocks output entirely rather than weakening it: when a
node has an \emph{irreversible} external effect whose identity the traces leave
\emph{under-determined} --- most sharply when the candidates are opposites, such
as adding versus removing --- the compiler emits no executable graph. Deferring
that choice to a runtime LLM node is explicitly not an acceptable resolution,
since a runtime model has strictly less evidence than the analysis had.

\paragraph{Vocabulary note.}
The skill's taxonomies are finer than the paper's prose. Its nine event-role
labels refine the seven categories of Section~\ref{sec:method-denoising}, and
its edge verdicts (\textsc{hard}, \textsc{transform}, \textsc{suspected},
\textsc{refuted}) do not map one-to-one onto the relation classes of
Section~\ref{sec:method-dependencies}; only \textsc{hard} and \textsc{suspected}
carry identical meaning in both. ``Conditional hard'' is represented in the
artifact as a hard edge on a guarded node.

\end{document}